\documentclass[twocolumn, prb, superscriptaddress, amsmath, amssymb,floatfix]{revtex4-2}
\usepackage{graphicx, nicefrac}
\usepackage[thinspace,thinqspace,amssymb]{SIunits}
\usepackage{tabto}
\usepackage[version=4]{mhchem}
\usepackage{chemformula}
\usepackage{color}
\usepackage{hyperref}
\usepackage{float}
\usepackage{ulem}
\usepackage{textgreek}

\renewcommand{\emph}{\textit}

\newcommand{\PETRA}{\mbox{PETRA~III}}

\newcommand{\STO}{SrTiO$_\text{3}$}

\begin{document}
\NumTabs{10}

\title{Spin polarization of the two-dimensional electron gas at the EuO/SrTiO$_3$ interface}

	\author{Paul Rosenberger}
	\affiliation{Fachbereich Physik, Universit\"at Konstanz, 78457 Konstanz, Germany}
	\affiliation{Institut f\"ur Physik, Johannes Gutenberg-Universit\"at, 55099 Mainz, Germany}

        \author{Andri Darmawan}
        \affiliation{Department of Physics and Center for Nanointegration (CENIDE), Universit\"at Duisburg-Essen, Lotharstr.~1, 47057 Duisburg, Germany}
	
	\author{Olena Fedchenko}
	\affiliation{Institut f\"ur Physik, Johannes Gutenberg-Universit\"at, 55099 Mainz, Germany}
	
	\author{Olena Tkach}
	\affiliation{Institut f\"ur Physik, Johannes Gutenberg-Universit\"at, 55099 Mainz, Germany}
	
	\author{Serhii V. Chernov}
	\affiliation{Stony Brook University, Department of Physics and Astronomy, Stony Brook, USA}
	
	\author{Dmytro Kutnyakhov}
	\affiliation{Deutsches Elektronen-Synchrotron DESY, 22607 Hamburg, Germany}

	\author{Moritz Hoesch}
	\affiliation{Deutsches Elektronen-Synchrotron DESY, 22607 Hamburg, Germany}

        \author{Markus Scholz}
	\affiliation{Deutsches Elektronen-Synchrotron DESY, 22607 Hamburg, Germany}

        \author{Kai Rossnagel}
        \affiliation{Institut f\"ur Experimentelle und Angewandte Physik, Christian-Albrechts-Universit\"at zu Kiel, 24098 Kiel, Germany}
	\affiliation{Ruprecht Haensel Laboratory, Deutsches Elektronen-Synchrotron DESY, 22607 Hamburg, Germany}
 
    \author{Rossitza Pentcheva}
    \affiliation{Department of Physics and Center for Nanointegration (CENIDE), Universit\"at Duisburg-Essen, Lotharstr.~1, 47057 Duisburg, Germany}
 
	\author{Gerd Sch\"onhense}
	\affiliation{Institut f\"ur Physik, Johannes Gutenberg-Universit\"at, 55099 Mainz, Germany}
	
	\author{Hans-Joachim Elmers}
	\affiliation{Institut f\"ur Physik, Johannes Gutenberg-Universit\"at, 55099 Mainz, Germany}

	\author{Martina M\"uller}
	\email{martina.mueller@uni-konstanz.de}
	\affiliation{Fachbereich Physik, Universit\"at Konstanz, 78457 Konstanz, Germany}

	\date{\today{}}

\begin{abstract}
Spin-polarized two-dimensional electron gases (2DEGs) are of particular interest for functional oxide electronics applications. The redox-created 2DEG residing on the strontium titanate, SrTiO$_3$ (STO), side of a europium monoxide (EuO)/SrTiO$_3$~(001) interface is expected to be significantly spin-polarized due to the proximity to the strong ($7\,\mu_B/f.u.$) Heisenberg ferromagnet EuO. We apply magnetic circular dichroism in the angular distribution (MCDAD) of photoemitted electrons to investigate whether and how the induced spin polarization of the 2DEG depends on the dimensionality of the overlaying EuO layer. The experimental data are complemented by density functional theory calculations with a Hubbard $U$ term (DFT+$U$). We show that the EuO/STO interfacial 2DEG is spin-polarized even for ultrathin EuO overlayers, starting at an EuO threshold thickness of only two monolayers. Additional EuO monolayers even increase the induced magnetic Ti moment and thus the spin polarization of the 2DEG. Our results and the potential to enhance the magnetic order of EuO by other proximity effects 
indicate that the EuO/STO~(001) interface is an ideal template for creating (multi-)functional spin-polarized 2DEGs for application in oxide electronics.
\end{abstract}

\maketitle
	
\section{Introduction}
\label{sec:introduction}
		
The observation of a conducting interface between the large-band gap insulators LaAlO$_3$ and SrTiO$_3$~\cite{Ohtomo2004} sparked the scientific interest in
oxide heterostructures with emergent properties. The two-dimensional electron gas (2DEG) near the interface can exhibit paramagnetism and ferromagnetism and even coexist with superconductivity~\cite{Brinkman2007magnetic,Bert2011}. 
The origin of the 2DEG~\cite{Li2020} is either associated with
a charge transfer at the interface to avoid diverging electric fields at the interface~\cite{Goniakowski2007,Hwang2006}, 
or to oxygen vacancies at the SrTiO$_3$ surface that arise during growth~\cite{Kalabukhov2007,Herranz2007}. The oxygen vacancies are present in crystalline oxide films that are grown
under reducing environments~\cite{Kormondy2015,Edmondson2018}. 
Oxygen vacancies in SrTiO$_3$ act as donors~\cite{Lin2013} and contribute charge carriers to the 2DEG. The corresponding states in the band gap were observed by angle-resolved photoemission spectroscopy~\cite{Kormondy2018,Zapf2022}. 
For functional oxide (spin-)electronic devices, it would be highly interesting to form a spin-polarized
2DEG at the interface. 
Spin-polarized 2DEGs at the interfaces of oxides have been observed for
LAO/BaTiO$_3$~\cite{Chen2018}, SrMnO$_3$/LaMnO$_3$~\cite{Nanda2008} and 
heterostructures with EuTiO$_3$~\cite{Gui2019,Lu2015,Stornaiuolo2015,DiCapuanpjQM2022,DiCapua2021}.

EuO/SrTiO$_3$ interfaces have attracted large interest~\cite{Kormondy2018,Gao2018,Lee2010,Wang2009,Lukashev2012,lomker_two-dimensional_2017}
because the conductivity mismatch in metal/semiconductor junctions can be overcome
by the ferromagnetic (FM) semiconductor EuO~\cite{Schmidt2000,caspers_electronic_2011,caspers_interface_2016}.
Therefore, the EuO/SrTiO$_3$ system could be used to fabricate
a spin metal–oxide–semiconductor field-effect transistor~\cite{Datta1990}. Recent studies show that the magnetic order of EuO can be enhanced and also designed by magnetic proximity coupling to magnetic transition metals of the 3$d$ group~\cite{rosenberger_proximity_2024,monkebuscher_modulation_2023}. This overcomes the relatively low Curie temperature of EuO ($69$~K) in the coupled 3$d$~FM/EuO bilayer, which behaves like a strong synthetic ferrimagnet.
Further, EuO exhibits a large ($0.6$~eV) exchange splitting in the conduction band~\cite{santos_determining_2008,mueller_exchange_2009,miao_magnetoresistance_2009,heider_temperature_2022}. If the vacancy-induced carriers are located at the EuO side of the interface, it could result in a $100$\% spin-polarized 2DEG ~\cite{jutong_electronic_2012}. 
However, previous experiments show that due to the
band alignment at the interface, the vacancy-induced carriers reside mostly within
SrTiO$_3$~\cite{Kormondy2018,Gao2018}. 
Thus, it is not yet clear whether the interface states formed at the 
EuO/SrTiO$_3$ interface are spin-polarized or not.
In this article, we show that the interface states are indeed spin-polarized.
We apply circular dichroism in the angular distribution of direct photoemitted electrons to demonstrate a spontaneous breaking of the time-reversal symmetry of the interface states, which is a clear signature of finite spin polarization of these states. Further insights on the electronic reconstruction and spin polarization of few monolayers (ML) EuO on SrTiO$_3$~(001) are obtained from density functional theory calculations with a Hubbard $U$ term (DFT+$U$).

\section{Experimental details}

\subsection{Sample preparation and handling}
\label{sec:preparation}
We have prepared EuO ultrathin films on TiO$_\text{2}$-terminated \STO:Nb substrates (Crystec GmbH) using molecular beam epitaxy (MBE) in ultrahigh vacuum (UHV). The base pressure of the oxide MBE system at University of Konstanz was \mbox{p$_\text{MBE}\,\leq\,2\,\times\,10^{-10}\,$mbar}. Prior to EuO growth, the substrates were annealed in UHV for $30\,$min at T$_\text{S}\,=\,500^\circ$C. EuO films of thicknesses of one, two and four monolayers were redox-grown as described elsewhere~\cite{lomker_two-dimensional_2017}. The substrate was kept at room temperature during deposition. 
Eu metal was evaporated from a Knudsen cell at a rate of r$_\text{Eu}\,=\,0{.}005\,$\AA/s. The rate was measured with a quartz crystal microbalance.
Subsequent \emph{in situ} X-ray photoelectron spectroscopy (XPS) was applied to confirm the stoichiometry of the deposited EuO film as well as the partial reduction of Ti$^{4+}$ in the substrate to Ti$^{3+}$, indicative for the formation of the interfacial two-dimensional electron gas~\cite{lomker_two-dimensional_2017}.\\
An ultra-high vacuum suitcase was used to transfer the samples \emph{in vacuo} to the beamline P04 at \PETRA~(DESY, Hamburg)~\cite{viefhaus_variable_2013}.\\
After transferring a sample to the liquid helium (LHe) cooled sample stage it was magnetized by approaching a permanent magnet. The measurement was then performed with the sample in the remanent magnetic state. 

\subsection{Setup}
\label{sec:SetupMethod}
For the photoemission measurements at beamline P04 (PETRA III, DESY Hamburg), photoelectrons were excited by circularly polarized soft X-rays.
We used a time-of-flight momentum microscope (MM) at the open port I of P04 with an energy resolution of $60$~meV at a sample temperature of 30~K~\cite{Medjanik2017,Fedchenko2024,Tkach2024}. 
The photoemission experiments were performed with the incidence angle of the photon beam at 22$^\circ$ with respect to the sample surface, and the azimuthal orientation of the sample has been adjusted so that the photon incidence plane coincides with the (100) axis of the \STO~substrate.
The coordinate system for the photoelectron momentum ($k_x,k_y,k_z$) was set to $k_z$ along the crystallographic [001] direction, i.e., surface normal, $k_x$ along [100] and $k_y$ along [010] in-plane directions, respectively.


\section{Results}
\label{sec:Result}
\subsection{\emph{In situ} sample characterization}

    \begin{figure*}[tbh]
        \centering
        \hspace*{-1mm}
        \includegraphics[width=1.80\columnwidth,clip]{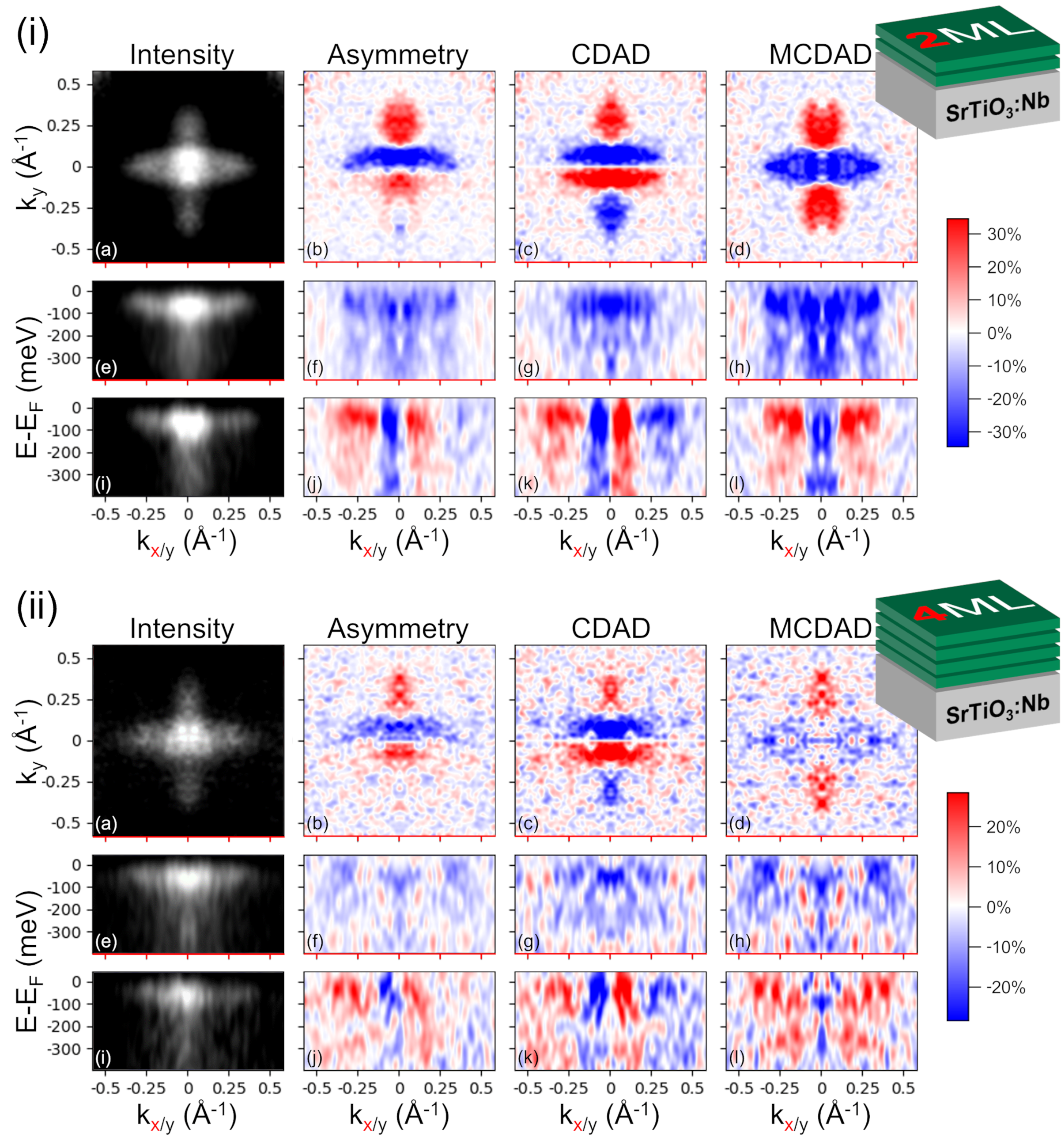}
        \caption{Photoemission maps obtained from the samples with (i) two, and (ii) four monolayers EuO on STO:Nb. Intensity, asymmetry, CDAD and MCDAD are shown (from left to right), each as functions of $k_x$ and $k_y$ at $E_\text{B}=50~{\rm meV}$~(top), $k_x$ and $E_\text{B}$ at $k_y=0$~(middle; for CDAD: $k_y \approx 0.02$~\AA$^{-1}$) as well as $k_y$ and $E_\text{B}$ at $k_x=0$~(bottom). All maps exhibit strong features: The intensity maps reveal the presence of the interfacial 2DEG. The asymmetry is pronounced and yields both a strong CDAD and MCDAD. The CDAD suggests that the bands extending along $k_x$ and $k_y$ are of different orbital character. The non-vanishing MCDAD indicates a finite spin polarization of the 2DEG.} 
        \label{fig:results:data_2ML}
        \label{fig:results:data_4ML}
    \end{figure*}

    \begin{figure*}[tbh]
	\centering
	\hspace*{-1mm}
	\includegraphics[width=1.80\columnwidth,clip]{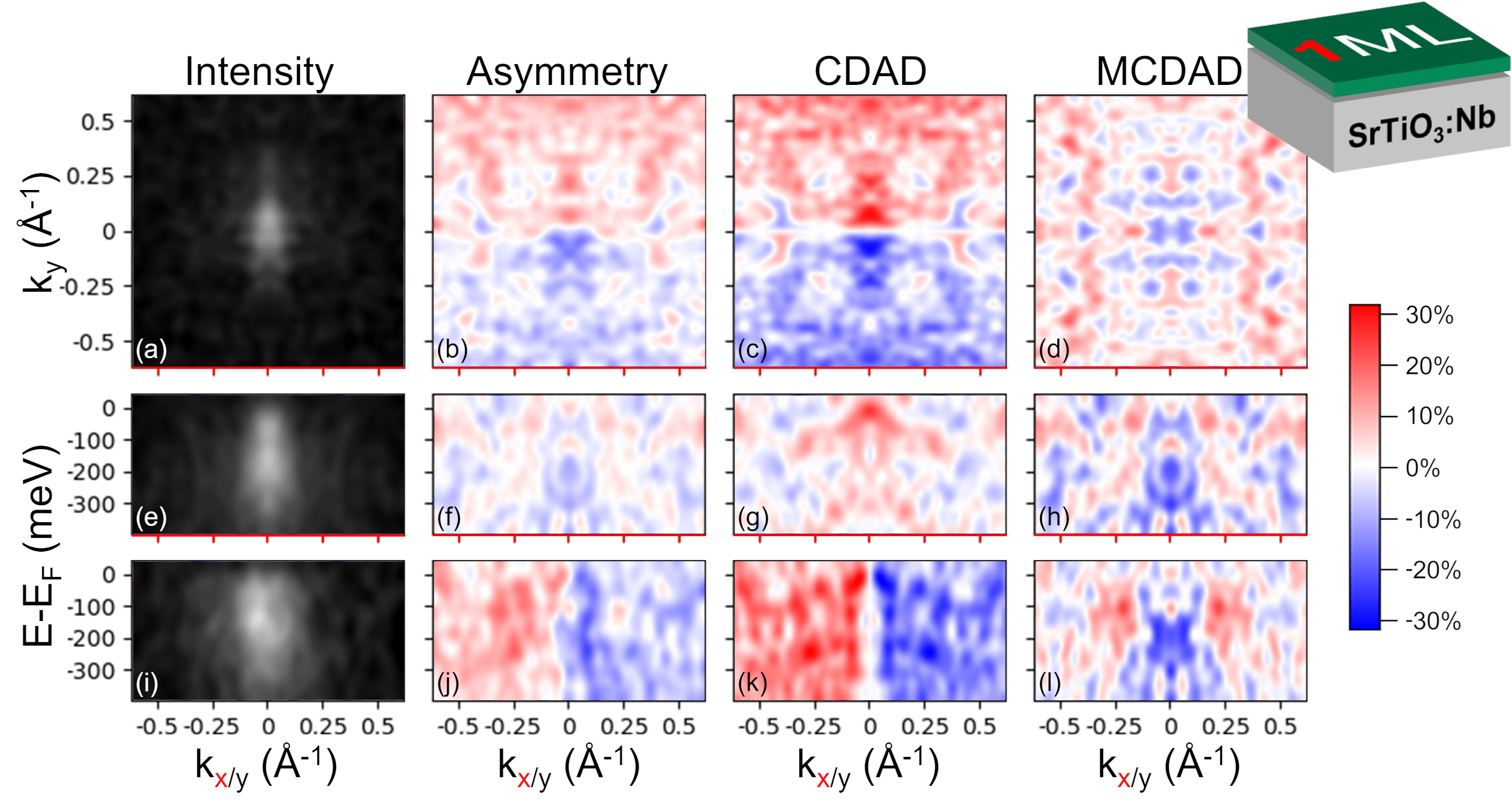}
	\caption{Photoemission maps obtained from the sample with one monolayer EuO on STO:Nb. Intensity, asymmetry, CDAD and MCDAD are shown (from left to right), each as functions of $k_x$ and $k_y$ at $E_\text{B}=50~{\rm meV}$~(top), $k_x$ and $E_\text{B}$ at $k_y=0$~(middle; for CDAD: $k_y \approx 0.02$~\AA$^{-1}$) as well as $k_y$ and $E_\text{B}$ at $k_x=0$~(bottom). The weak features in the intensity maps suggest that the redox-formed interfacial 2DEG is only weak. The asymmetry pattern is antisymmetric with respect to $k_y=0$. While the CDAD exhibits clear features that, however, differ from those of the other samples, the MCDAD pattern is irregular, reflecting either the limited SNR or a different electronic structure than that of the other samples.}
	\label{fig:results:data_1ML}
	\end{figure*}

    \begin{figure}[t]
	\hspace*{-1mm}
	\includegraphics[width=0.95\columnwidth]{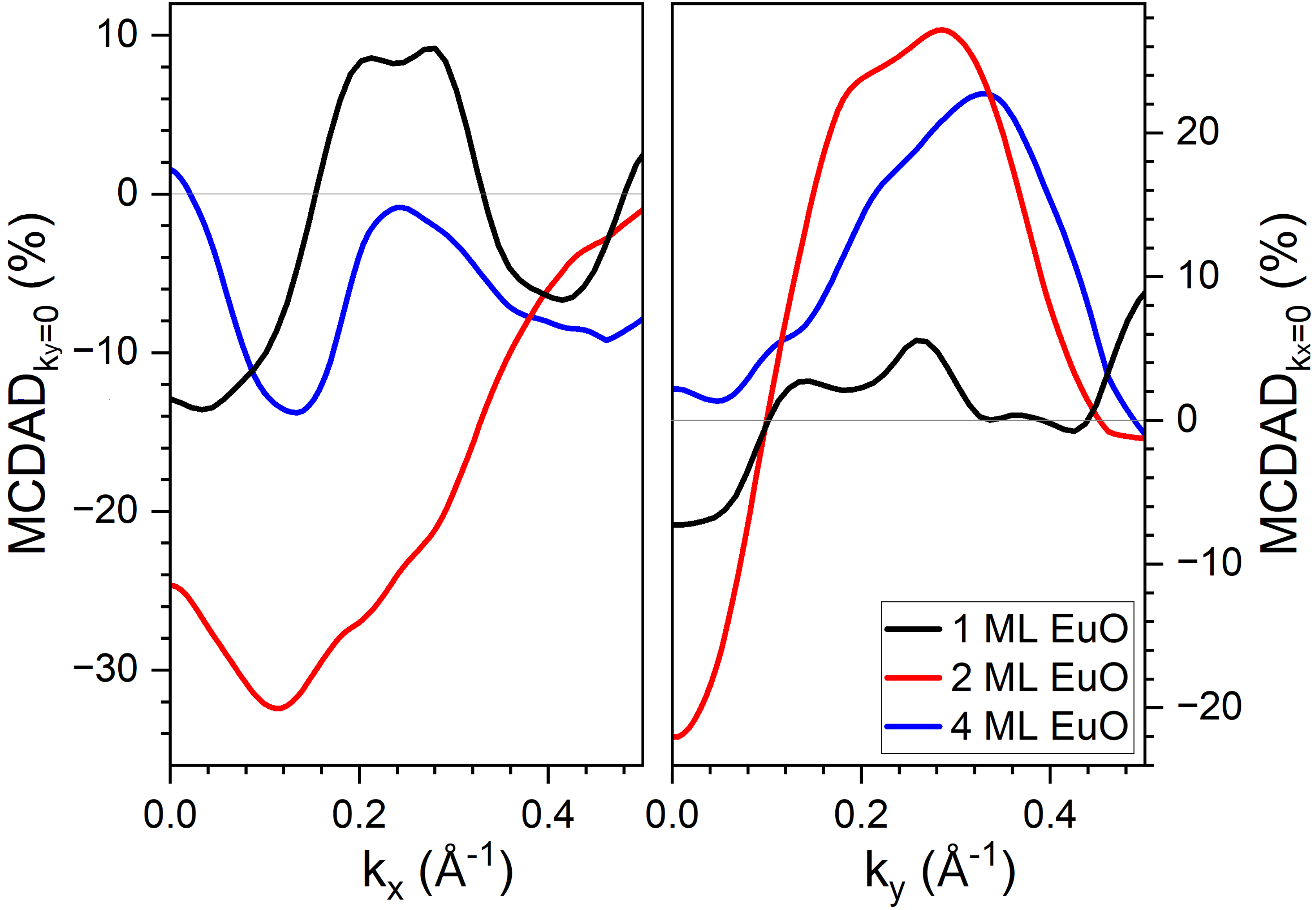}
	\caption{MDCs along $k_x$ (left) and $k_y$ (right) of the three samples displaying the averaged MCDAD of $-0.06$~\AA$^{-1}<k_y<+0.06$~\AA$^{-1}$ and $-0.06$~\AA$^{-1}<k_x<+0.06$~\AA$^{-1}$, respectively, at $E_\text{B}=50{\rm meV}$. The data are smoothed by locally weighted scatterplot smoothing over a span of $\approx 0.24$~\AA$^{-1}$.}
	\label{fig:results:MDC_MCDAD}
    \end{figure}

For all three samples, XPS of the Ti~2$p$ core level revealed the formation of Ti$^{3+}$ upon Eu metal deposition, indicating redox-driven formation of a 2DEG at the metal oxide/STO interfaces~\cite{lomker_two-dimensional_2017}.
From XPS of the Eu~3$d_\text{5/2}$ core level we find that EuO is formed with contributions of metallic Eu. The proportion of metallic Eu increases with increasing EuO thickness. This trend was expected because the oxygen ion conductivity of a material not only increases with increasing temperature but is also an important driving parameter for the interfacial redox reaction~\cite{lomker_two-dimensional_2017,rosenberger_europium_2022,lomker_redox-controlled_2019,mueller_hard_2022}. We consider the off-stoichiometry an advantage. Eu(O) oxidizes even when stored under UHV conditions~\cite{rosenberger_emerging_2024}. Hence, a sample that is perfectly stoichiometric immediately after preparation in the lab will contain some undesired Eu$^{3+}$ when investigated later at the synchrotron. In contrast, in a sample that initially contains Eu metal, Eu$^0$ will oxidize to Eu$^{2+}$, i.e., EuO, before the sample starts to degrade by overoxidation. Therefore, our samples do not degrade but ``ripen'' in the storage time between preparation in the lab and measurement at P04.
LEED and RHEED indicate that the redox-grown EuO films are amorphous, in agreement with a previous study~\cite{lomker_two-dimensional_2017}.

\subsection{Momentum microscopy}
To investigate the proximity-induced magnetic order of the 2DEG at the EuO/STO interface and its dependence on the EuO thickness, we performed ToF-MM measurements of the valence band using left and right circularly polarized light (intensities $I^-$ and $I^+$, respectively). The photon energy was set to $h\nu=466$~eV, i.e., to the Ti~L$_\text{2}$ resonance, to enhance the signal from the buried interface~\cite{Berner2013}.

Figures~\ref{fig:results:data_2ML}~(i), (ii) and \ref{fig:results:data_1ML} show maps of the photoemission intensities and asymmetries of the samples with EuO thicknesses of $2$~ML, $4$~ML and $1$~ML, respectively. In each figure, from left to right these are the sum intensity as a function of binding energy $E_\text{B}$ and parallel momentum $k_x$ and $k_y$,
\begin{equation}
I_{\rm sum}(E_\text{B},k_x,k_y) = I^+(E_\text{B},k_x,k_y) + I^-(E_\text{B},k_x,k_y),
\end{equation}
the difference,
\begin{equation}
A(E_\text{B},k_x,k_y) = I^+(E_\text{B},k_x,k_y) - I^-(E_\text{B},k_x,k_y),
\end{equation}
the circular dichroism in the angular distribution (CDAD)~\cite{Westphal1989},
\begin{multline}
{\rm CDAD}(E_\text{B},k_x,k_y) = \\ A^+(E_\text{B},k_x,k_y) - A^-(E_\text{B},k_x,-k_y),
\end{multline}
and the magnetic circular dichroism in the angular distribution (MCDAD)~\cite{Schneider1991,Kuch2001},
\begin{multline}
{\rm MCDAD}(E_\text{B},k_x,k_y) = \\ A^+(E_\text{B},k_x,k_y) + A^-(E_\text{B},k_x,-k_y).
\end{multline}

In the figures, the following maps of these measures are shown:

\noindent top: \tab\tab\tab $N(E_\text{B}=50~{\rm meV},k_x,k_y)$,\\ 
middle: \tab\tab $N(E_\text{B},k_x,k_y=0)$,\\ 
bottom: \tab\tab $N(E_\text{B},k_x=0,k_y)$,\\ 

where $N$ denotes $I$, $A$, CDAD, and MCDAD, respectively. 
Note that by definition the CDAD at $k_y=0$ vanishes. Therefore, in Fig.~\ref{fig:results:data_1ML}~(g), we present the section $k_y \approx 0.02$~\AA$^{-1}$. The data were two-fold symmetrized, i.e. in our representations it is $N\,(k_x)=N\,(-k_x)$.

We first discuss the results for the ferromagnetic $2$~ML and $4$~ML EuO on STO:Nb sample.
For the samples with $2$~ML and $4$~ML EuO on STO:Nb, a blurred, yet well-resolved (band) structure can be observed in the sum intensity images, Figs.~\ref{fig:results:data_2ML}~(i)~(a,e,i) and Figs.~\ref{fig:results:data_4ML}~(ii)~(a,e,i). The subfigures (a) show the Fermi surfaces around the $\Gamma$-point. Subfigures (e) and (i) are the corresponding energy-momentum maps revealing the electronic band structure along $k_x$ and $k_y$, respectively. They reveal parabolic structures extending from $E_F$ to the band bottoms at approximately $-150$~meV and $-400$~meV, respectively. The corresponding widths of the structures along $k$ are $\approx 0{.}75$~\AA$^{-1}$ and $\approx 0{.}3$~\AA$^{-1}$. These widths match well with those estimated from an earlier angle-resolved photoelectron spectroscopy~(ARPES) study of the EuO/STO 2DEG for the Ti~3$d_{xz/yz}$ and Ti~3$d_{xy}$ subbands~\cite{lomker_two-dimensional_2017}. We therefore state that our sum intensity images, Figs.~\ref{fig:results:data_2ML}~(i)~(a,e,i) and Figs.~\ref{fig:results:data_4ML}~(ii)~(a,e,i), confirm that the interfacial 2DEG was formed.\\
For both samples, a strong difference intensity $A$ is observed [see
Figs.~\ref{fig:results:data_2ML}~(i)~(b,f,j)] showing a similar pattern.
The difference is larger for the $2$~ML sample compared to the $4$~ML sample.
This can be explained by the limited inelastic mean free path of the photoexcited electrons. The thicker EuO layer scatters photoexcited electrons, destroying the momentum information they carry.

The CDAD calculated from $A$ [Figs.~\ref{fig:results:data_2ML}~(i)~(c,g,k) and Figs.~\ref{fig:results:data_4ML}~(ii)~(c,g,k)] is antisymmetric with respect to the $k_x$-axis as forced by definition. The band extending along the $k_x$-axis close to $k_y=0$ has the opposite sign compared to the band along the $k_y$-axis, indicating a different orbital character of these two bands. Similar patterns appear for both the $2$~ML and the $4$~ML sample.

The MCDAD shown in Figs.~\ref{fig:results:data_2ML}~(i)~(d,h,i) and Figs.~\ref{fig:results:data_4ML}~(ii)~(d,h,i)
reveals non-zero values in the region of the interface bands with a similar pattern for the $2$~ML and $4$~ML sample. The non-vanishing MCDAD indicates a time-reversal symmetry breaking that can only be explained by a non-vanishing spin polarization of the interface bands. 
The sign of the MCDAD does not necessarily reflect the sign of the spin polarization, as it results from a combination of exchange splitting and spin-orbit interaction~\cite{Kuch2001}.

For a quantitative comparison, the momentum distribution curves (MDCs) along the $k_x$- and $k_y$-axes
are shown in Fig.~\ref{fig:results:MDC_MCDAD}. 
To increase the signal-to-noise ratio (SNR) the MDCs are averaged by $\pm 0.06$~\AA$^{-1}$
along the perpendicular momentum axis  and by $E_\text{B}>0.1$~eV along the binding energy axis.
The maximum MCDAD signal amounts to $\pm 0.3~I_{\rm max}$ for the $2$~ML sample and
$\pm 0.2~I_{\rm max}$ for the $4$~ML sample.

The single monolayer of EuO on STO:Nb shows a different behavior.
The photoemission intensity is comparatively low 
[Fig.~\ref{fig:results:data_1ML}~(a,e,i)] and does not show the cross-shaped pattern observed for the $2$~ML and $4$~ML sample.
Thus, the 2DEG at the EuO/STO interface is weak.
The difference $A$ shown in Fig.~\ref{fig:results:data_1ML}~(b)
essentially shows an antisymmetric pattern with respect to the $k_x$-axis.
The corresponding CDAD [Fig.~\ref{fig:results:data_1ML}~(b,f,j)] clearly shows plus/minus features but appears distinct from the CDAD observed for the $2$~ML and $4$~ML samples.
It is positive for $k_y>0$ and negative for $k_y<0$.

The MCDAD [Fig.~\ref{fig:results:data_1ML}~(b,f,j)] shows an irregular pattern which might be explained by the limited SNR or it reflects a different electronic structure.
Only the dispersion curve along the $k_y$-axis shown in Fig.~\ref{fig:results:data_1ML}~(l) exhibits a pattern that resembles that of the $2$ and $4$~ML samples. In contrast, the MDC shown in Fig.~\ref{fig:results:MDC_MCDAD} is sizable only along the $k_x$-axis. However, it differs from the MDCs of $2$~ML and $4$~ML, supporting our assumption that the features obtained for a single monolayer are due to an insufficient SNR or of different physical origin than those observed for $2$~ML and $4$~ML EuO/STO.

\section{Theoretical modeling}

\begin{figure}[t]
        \hspace*{-1mm}
	\includegraphics[width=\columnwidth]{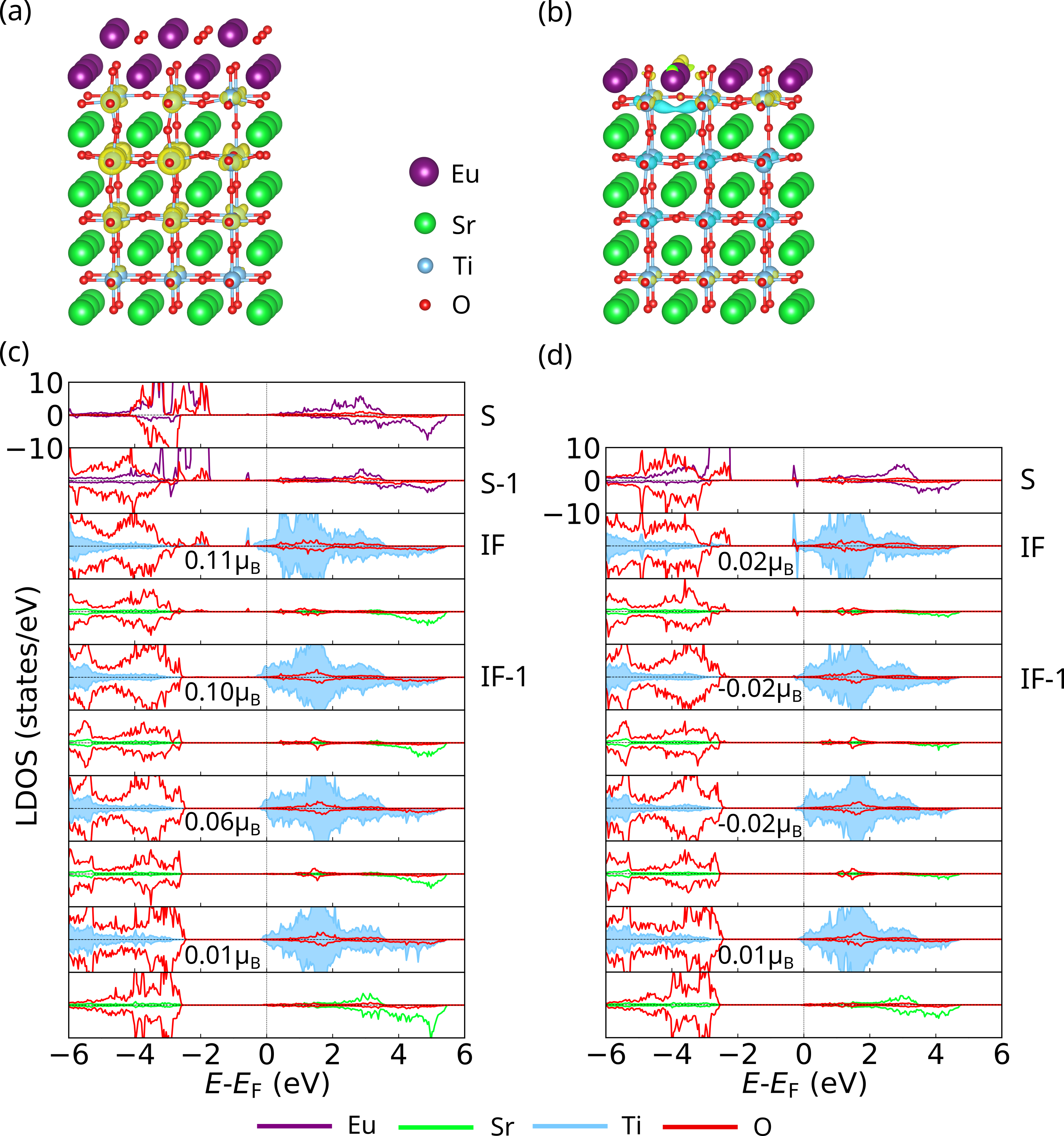}
	\caption{Side view of the relaxed structure and spin density of (a) $2$~ML EuO and (b) $1$~ML EuO on STO~(001) with an oxygen vacancy in the topmost TiO$_2$ layer and the second vacancy in the subsurface SrO layer. The spin density of 1-2ML EuO on STO~(001) is integrated between $-0.4$~eV and the Fermi level. Yellow and cyan areas indicate positive and negative spin density, respectively. Layer-, element-, and spin-resolved density of states of (c) $2$~ML EuO on STO~(001) and (d) $1$~ML EuO on STO~(001) together with the average of Ti magnetic moment for each TiO$_2$ layer in $\mu_{\rm B}$.}
	\label{fig:spinldos}
    \end{figure}

We performed DFT+$U$ calculations to model EuO/SrTiO$_3$(001)  using the Vienna $ab\;initio$ simulation package (VASP)~\cite{Kresse1993,Kresse1996} with the projector augmented wave (PAW)~\cite{Bloch1994,Kresse1999} basis and the PBEsol approximation for the exchange correlation functional~\cite{Perdew2008}. The DFT+$U$ approach within the Dudarev's scheme~\cite{Dudarev1998} was used with an on-site Hubbard $U$ term of $2$~eV and $7.5$~eV for the Ti~3$d$ and Eu~4$f$ states, respectively. The latter value leads to a very good agreement with angle-integrated valence band spectra~\cite{DiCapua2021} concerning the position of the localized Eu~4$f$ states. The system consists of $1$ or $2$~ML of EuO (rotated by $45^{\circ}$)  on top of STO~(001) where the Eu ion is located at a tentative Sr site on top of the TiO$_2$ layer that was reported previously to be  the most stable configuration~\cite{Gao2018,Kormondy2018}. We employed a $2a \times 3a \times 4a$ STO slab with two oxygen vacancies  including $20$~$\text{\AA}$ vacuum region to prevent any interaction between the slab and its periodic images. The first oxygen vacancy is located in the topmost TiO$_2$ layer and the second is in the SrO subsurface layer with a distance of $6.2$~\AA \, [Figs.~\ref{fig:spinldos}(a) and ~\ref{fig:spinldos}(b)].  This configuration was found previously to be most suitable to model the 2DEG at the STO~(001) \cite{Jeschke2015} and ETO(001)~\cite{DiCapua2021}. The lateral lattice constant is fixed to the experimental value of STO, $3.905$~\AA, while the calculated bulk EuO lattice constant is $5.18$~\AA. Thus the EuO layer is subject to  $6.6$\% tensile strain. We used a plane-wave cutoff energy of $400$~eV and sampled the Brillouin zone with $6\times4 \times 1$ Monkhorst-pack $k$-points. The ionic positions are fully relaxed until the forces were less than $0.01$~eV/\AA.

A side view of the relaxed structure and  spin density and the layer-resolved density of states (LDOS) of $2$ and $1$~ML of EuO on STO~(001) is shown in Fig.~\ref{fig:spinldos}. Both the spin-density and LDOS indicate the formation of a spin-polarized 2DEG in particular for the $2$~ML EuO on STO~(001): the  averaged Ti magnetic moment in the topmost TiO$_2$ layer  is  $\sim0.11$~$\mu_{\rm B}$/Ti for $2$~ML EuO on STO and $0.02~\mu_B$/Ti for $1$~ML Eu on STO. For $3$~ML EuO (not shown), the Ti magnetic moment is even larger than for $2$~ML, i.e. $\sim0.16$~$\mu_{\rm B}$/Ti. While the localized Eu~4$f$ states lie below $-2$~eV, the Ti~3$d$ states show a localized in-gap state at $-0.5$~eV in the majority spin channel which hybridizes with Eu~5$d$ states. The in-gap state is aligned ferromagnetically with the Eu~4$f$ states due to the proximity effect. Furthermore, dispersive spin-polarized contributions to the 2DEG  between $-0.3$~eV and $E_{\rm F}$ are observed both in the interface and the neighboring deeper layer. In contrast, the spin polarization is nearly quenched in the case of $1$~ML EuO  on STO~(001), the localized in-gap states lying closer to the Fermi level appear in both spin channels, thus cancel out.

    \begin{figure}[t]
	\hspace*{-1mm}
	\includegraphics[width=\columnwidth]{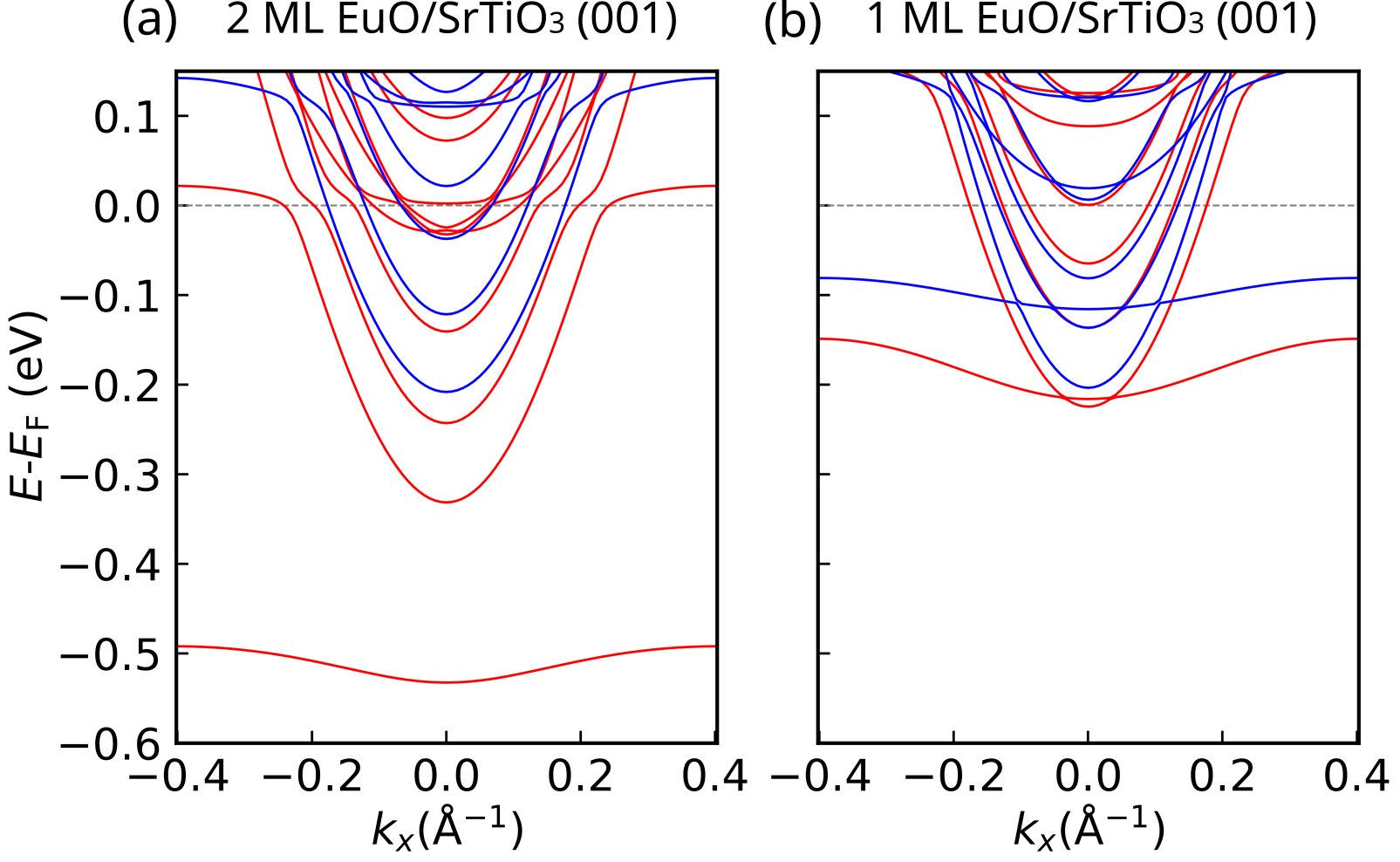}
	\caption{DFT+$U$ spin resolved electronic band structure of (a) $2$~ML EuO on STO~(001) and (b) $1$~ML EuO on STO~(001) along $\Gamma-X$ direction. Red/blue denotes majority/minority bands.}
	\label{fig:bandstructure}
    \end{figure}
 
The spin-resolved band structure of $2$ and $1$~ML EuO on STO~(001) presented  in Fig.~\ref{fig:bandstructure}~(a) and (b), shows significant differences in the two cases. The dispersive majority bands of $d_{xy}$ character reach much deeper below the Fermi level (up to $-0.35$~eV) for $2$~ML EuO on STO~(001) than for $1$~ML EuO on STO~(001) ($-0.23$~eV). Moreover, the exchange splitting is one order of magnitude larger in the former case. In the latter case the localized band in the majority channel lies $0.1$~eV lower than in the minority channel. Overall, the band structure confirms the much more robust spin polarization for $2$~ML than for $1$~ML EuO on STO~(001), consistent with experiment.

\section{Discussion}
\label{sec:discussion}
Our present study addresses the spin polarization of the redox-created 2DEG at the EuO/STO~(001) interface by means of circular dichroism in ToF-MM and DFT+$U$ calculations. Both experiment and theory reveal that an EuO film as thin as $2$~ML induces a magnetic moment in the thus spin-polarized 2DEG residing at the STO side of the interface. The induced magnetic moment is ferromagnetically aligned with the magnetization of the EuO overlayer. In contrast, no sizeable MCDAD signal is observed for a single EuO ML, indicating that spin polarization is absent. This is in line with the DFT+$U$ calculations, suggesting that the induced Ti magnetic moment for a system with a single EuO monolayer is less than $20$\% of the magnetic moment induced by $2$~ML EuO. This is supported by previous SQUID (superconducting quantum interference device) magnetometry measurements that find that two monolayers EuO on STO are ferromagnetic, while a single monolayer remains paramagnetic~\cite{lomker_two-dimensional_2017}.\\
Increasing the EuO thickness to $4$~ML results in the same characteristics in the photoemission data as observed for $2$~ML although with lower intensity due to stronger scattering of the photoelectrons from the interface in the now thicker EuO layer. Experimental evidence for a finite spin polarization of the EuO/STO interfacial 2DEG for EuO overlayer thicknesses larger than $2$~ML and calculations predicting an even increased Ti magnetic moment for three instead of two monolayers of EuO strongly suggest that the threshold EuO thickness for obtaining spin polarization in the redox-created 2DEG at the EuO/STO interface is as low as $2$~ML.\\
We assume that at finite temperature the induced Ti magnetic moment scales with the size of the Eu magnetic moment. In a very recent work we showed that a magnetic proximity effect at 3$d$~ferromagnet/EuO interfaces enhances the magnetic order in EuO effectively in the 2D limit~\cite{rosenberger_proximity_2024}. We therefore believe that the degree of spin polarization of the 2DEG at the EuO/STO interface can be significantly enhanced by depositing a 3$d$~ferromagnet overlayer on top of the few monolayer thin EuO on STO.\\

In summary, we studied the spin polarization of the redox-created 2DEG at the EuO/STO~(001) interface by circular dichroism in ToF-MM and DFT+$U$. We found that the 2DEG is significantly spin-polarized due to the proximity to the ferromagnetic EuO overlayer. The threshold EuO thickness for inducing spin polarization is as low as $2$~ML. Since the magnetic order of EuO can be enhanced further by other proximity effects,  the EuO/STO (001) interface turns out to serve as an ideal template for creating (multi-)functional spin-polarized 2DEGs for application in oxide electronics.

\begin{acknowledgments}
    This work was supported by the Deutsche Forschungsgemeinschaft through Sonderforschungsbereich SFB 1432 (Project No. 425217212, Subproject No. B03) and Transregio TRR 173 (Project No. 268565370, Subproject No. A02)
    and by the German Federal Ministry of Education and Research under framework program ErUM (project 05K22VL1 and 05K22UM2). We acknowledge DESY (Hamburg, Germany), a member of the Helmholtz Association HGF, for the provision of experimental facilities. Parts of this research were carried out at PETRA III using beamline P04. Beamtime was allocated for proposal I-20220420. R.P. and A.D. acknowledge support by the German Research Foundation (DFG, Deutsche Forschungsgemeinschaft) within the Collaborative Research Centre CRC 1242 (Project No. 278162697, Subproject No. C02) and computational time at Leibniz Rechenzentrum, project pr87ro.
\end{acknowledgments}

%

\end{document}